\documentstyle[11pt,oneside,amssymb,array,amstex]{amsart}

\newcommand{\al}{\alpha}
\newcommand{\be}{\beta}
\newcommand{\pa}{\partial}

\newcommand{\om}{\omega}
\newcommand{\de}{\delta}

\newcommand{\rar}{\rightarrow}

\def\fun#1#2{\lower3.6pt\vbox{\baselineskip0pt\lineskip.9pt
  \ialign{$\mathsurround=0pt#1\hfil##\hfil$\crcr#2\crcr\sim\crcr}}}
\newskip\humongous \humongous=0pt plus 1000pt minus 1000pt

\newif\ifdtup

\begin{document}

\begin{titlepage}

\begin{flushright}
M\'exico ICN-UNAM 99-01
\\January 23, 1999
\end{flushright}

\vskip 1.6cm

\begin{center}

{\Large Different faces of harmonic oscillator}

\vskip 0.6cm

{\it Alexander Turbiner}$^{\dagger}$
\vskip 0.5cm
Instituto de Ciencias Nucleares, UNAM, Apartado Postal 70--543,
\\04510 Mexico D.F., Mexico

\end{center}

\vskip 2.cm

\centerline{Abstract}

\begin{quote}
Harmonic oscillator in Fock space is defined. Isospectral as well
as polynomiality-of-eigenfunctions preserving,
translation-invariant discretization of the harmonic oscillator
is presented. Dilatation-invariant and
polynomiality-of-eigenfunctions preserving discretization is also
given.
\end{quote}
\vskip 2.5cm
\begin{center}
Contribution to the Proceedings of the Conference {\it SIDE III},
Roma, Italy, May 1998\\ to be published by CRM Press, Canada
\end{center}

\vfill
\noindent
$^\dagger$On leave of absence from the Institute for Theoretical
and Experimental Physics, Moscow 117259, Russia\\ E-mail:
turbiner@@xochitl.nuclecu.unam.mx

\end{titlepage}
Undoubtedly the harmonic oscillator plays a fundamental role in
science. Goal of present note is to give a review of different
representations of quantum harmonic oscillator and its
deformations in terms of the elements of Heisenberg algebra,
differential, finite-difference, discrete operators.

The Hamiltonian of harmonic oscillator is defined by
\begin{equation}
\label{e1.1}
        {\cal H}= -\frac{\pa^{2}}{\pa x^{2}} +
        \om^2 x^{2} \ ,
\end{equation}
where $\om$ is the oscillator frequency. The eigenfunctions and
eigenvalues are given by
\begin{equation}
\label{e1.2}
\Psi_{k}(x) = H_k(\sqrt{\om}x) e^{-\om\frac{x^{2}}{2}}\ ,\ E_k=\om (2k +1),\ k=0,1,\ldots
\end{equation}
where $H_k$ is the $k$th Hermite polynomial in standard notation.
Without a loss of generality all normalization constants we put
equal to 1. The Hamiltonian (\ref{e1.1}) is $Z_2$-invariant, $x
\rar -x$, which leads to two families of eigenstates: even and
odd, symmetric and anti-symmetric with respect of reflection,
correspondingly. This property is coded in parity of the Hermite
polynomials:
\begin{equation}
\label{e1.3}
 H_{2n+p}(\sqrt{\om}x)= x^p L_n^{(p-\frac{1}{2})}(\om x^{2})\ ,\ n=0,1,\ldots
\end{equation}
where $L_n^{(\al)}(y)$ is the $n$th associated Laguerre
polynomial in standard notation, and $p=0,1$ has a meaning of
parity. Hereafter we can call a ground state the lowest energy
state of parity $p$:
\begin{equation}
\label{e1.4}
\Psi_{0}^{(p)}(x) = x^p e^{-\om\frac{x^{2}}{2}}\ ,
\end{equation}
Thus the formula (4) makes an unification of both possible values
of parity and for the sake of simplicity we will call (4) the
ground state eigenfunctions without specifying parity.

Make a gauge rotation of the Hamiltonian (1) taking a gauge
factor the ground state eigenfunction (4) and change variable $x$
to $y=\om x^2$, which incorporate the reflection symmetry.
Finally, after dropping off the constant terms we get an operator
\begin{equation}
\label{e1.5}
h(y,\pa_y)\ =\ -\frac{1}{\om}\ (\Psi_0^{(p)}(x))^{-1}{\cal
H}\Psi_0^{(p)}(x)\mid_{y=\om x^2}\ =\ 4y\pa_y^2 -
4(y-p-\frac{1}{2})\pa_y\
\end{equation}
with the spectrum $(-4n)$, where $n=0,1,2\ldots$. The operator
(5) simultaneously describes a family of eigenstates of positive
parity if $p=0$ and a family of eigenstates of negative parity if
$p=1$. We will call (5) {\it the algebraic form} of the
Hamiltonian of the harmonic oscillator. The word `algebraic'
reflects the fact that the operator (5) has a form of linear
differential operator with polynomial coefficients and
furthermore possesses infinitely-many polynomial eigenfunctions.
The latter implies that any eigenfunction can be found by
algebraic means by solving a system of linear algebraic equations

The algebraic form (5) admits a generalization of the original
Hamiltonian (1) we started with. If we assume that the parameter
$p$ can take any real value, $p>-1/2$, one can make a inverse
gauge transformation of the operator (5) back to the Hamiltonian
form and we arrive at
\[
{\om}\ y^{p/2}e^{- y/2}\bigg[4y\pa_y^2 - 4(
y-p-\frac{1}{2})\pa_y\bigg]y^{-p/2}e^{y/2}\mid_{x=\sqrt{
\frac{y}{\om}}}
\]
\begin{equation}
\label{e.1.6}
=\bigg[\pa_x^2 - \om^2 x^2
-\frac{p(p-1)}{x^2}\bigg]\equiv -{\cal H}_k\ ,
\end{equation}
which is known in literature as Kratzer Hamiltonian. It is worth
to mention that this Hamiltonian coincides also with 2-body
Calogero Hamiltonian. Hereafter we will call the system
characterized by the Hamiltonian (6) {\it the harmonic
oscillator}.

The resulting Hamiltonian (6) is characterized by the
eigenfunctions
\begin{equation}
\label{e1.7}
\Psi_{k}(x) = x^p L_n^{-\frac{1}{2}+p}(\om x^{2})
e^{-\om\frac{x^{2}}{2}}\ ,
\end{equation}
which coincides with (2) at $p=0,1$. The spectrum (6) is still
equidistant with energy gap $\om$ and after appropriate shift of
the reference point it coincides with the spectrum of the
original harmonic oscillator (1). Thus, the deformation of (1) to
(6) is isospectral, which is, of course, well-known.

In order to move ahead let us introduce a notion of the Fock
space. Take two operators $a$ and $b$ obeying the commutation
relation
\begin{equation}
\label{e.2.1}
             [a,b] \equiv ab  -  ba \ =\ I,
\end{equation}
with the identity operator $I$ on the r.h.s. -- they span a
three-dimensional Lie algebra which is called the Heisenberg
algebra $h_3$. By definition the universal enveloping algebra of
$h_3$ is the algebra of all normal-ordered polynomials in $a,b$:
any monomial is taken to be of the form $b^k a^m$
\footnote{Sometimes this is called the Heisenberg-Weyl algebra}.
If, besides the polynomials, all entire functions in $a,b$ are
considered, then the {\it extended} universal enveloping algebra
of the Heisenberg algebra appears or in other words, the extended
Heisenberg-Weyl algebra. In the (extended) Heisenberg-Weyl
algebra one can find the non-trivial embedding of the Heisenberg
algebra: non-trivial elements obeying the commutation relations
(\ref{e.2.1}), whose can be treated as a certain type of quantum
canonical transformations. We say that the {\it (extended) Fock
space, $\cal F$} is determined if we take the (extended)
universal enveloping algebra of the Heisenberg algebra and attach
to it the vacuum state $|0>$ such that
\begin{equation}
\label{e.2.2}
a|0>\  = \ 0\ .
\end{equation}
It is easy to check that the following statement holds if the
operators $a,b$ obey (\ref{e.2.1}), then the operators
\[
J^+_n = b^2 a - n b\ ,
\]
\begin{equation}
\label{e.2.3}
J^0_n = ba - {n \over 2}\ ,
\end{equation}
\[
J^-_n=a\ ,
\]
span the $sl_2$-algebra with the commutation relations:
\[
[J^0,J^{\pm}]=\pm J^{\pm}\ ,\  [J^+,J^-]=-2J^0\ ,
\]
where $n \in {\bf C}$
\footnote{For details and discussion see, for example,
\cite{Turbiner:1997}}. For the realization (\ref{e.2.3}) the
quadratic Casimir operator is equal to
\begin{equation}
\label{e.2.4}
C_2 \equiv \frac{1}{2}\{J^+_n,J^-_n\} - J^0_n J^0_n =
-\frac{n}{2} \bigg(\frac{n}{2} + 1\bigg)\ ,
\end{equation}
where $\{,\}$ denotes the anticommutator and is $c-$number. If $n
\in
\mathbb{Z}_+$, then (\ref{e.2.3}) possesses a finite-dimensional,
irreducible representation in Fock space leaving invariant the
space of polynomials in $b$:
\begin{equation}
\label{e.2.5}
{\cal P}_{n}(b) \ = \ \langle 1, b, b^2, \dots , b^n \rangle
|0\rangle,
\end{equation}
of dimension $\dim{\cal P}_{n}=(n+1)$. The spaces ${\cal P}_n$
possess a property that ${\cal P}_n \subset {\cal P}_{n+1}$ for
each $n \in
\mathbb{Z}_+$ and form an infinite flag and
$$\bigcup_{n \in \mathbb{Z}_+} {\cal P}_n = \mathcal{P}.$$
It is evident that any polynomial in generators $J^{0,-}_n$
operator preserves the flag of $\mathcal P.$ Such an operator we
will call {\it $sl_2$-exactly-solvable operator}.

Take as an example the $sl_2$-exactly-solvable operator of the
form
\begin{equation}
\label{e.2.6}
h_f(b,a)\ =\ 4J^0 J^- - 4 J^0 + 4(p+\frac{1}{2}) J^- \ =\ 4ba^2 -
4(b-p-\frac{1}{2})a\ ,
\end{equation}
where $p$ is a parameter and $J^{\pm,0}\equiv J^{\pm,0}_0$ (see
(10)). One can demonstrate that the eigenfunctions of $h_f$ are
the associated Laguerre polynomials of the argument $b$,
$L_n^{(p-\frac{1}{2})} (b)$ and their eigenvalues, $E_n=-4 n$.

As the next step we consider two different realization of the
Heisenberg algebra (\ref{e.2.1}) in terms of differential and
finite-difference operators. A traditional realization of
(\ref{e.2.1}) appearing in all text-books is the
coordinate-momentum representation:
\begin{equation}
\label{e.3.1}
a\ =\ \frac{d}{dy} \equiv \pa_y\ ,\ b\ =\ y\ ,
\end{equation}
where the operator $b=y$ stands for the multiplication operator
on $y$ in a space of functions $f(y)$. In this case the vacuum is
a constant and without a loss of generality we put $|0>\  = \ 1$.
Recently, a finite-difference analogue of (\ref{e.3.1}) has been
found \cite{Smirnov:1995}:
\begin{equation}
\label{e.3.2}
a\ =\ {\cal D}_+ ,\quad b\ =\ y(1-\de{\cal D}_-) \ ,
\end{equation}
where
\[
{\cal D}_{\pm} f(y) = \frac{f(y\pm\de) - f(y)}{\pm\de}\ ,
\]
is the finite-difference operator, $\de$ is real number and
${\cal D}_{\pm}(-\de) = {\cal D}_{\mp}(\de)$. A remarkable
property of this realization is that the vacuum remains the same
for both cases (\ref{e.3.1})-(\ref{e.3.2}) and it can be written
as $|0>\ =\ 1$.

Substitution of (\ref{e.3.1}) into (\ref{e.2.6}) leads to the
operator (5) -- the algebraic form of the Hamiltonian of the
harmonic oscillator. Thus, the operator (\ref{e.2.6}) can be
called the {\it algebraic form of the Hamiltonian of the harmonic
oscillator in the Fock space}. It is evident that the procedure
of realization of the Heisenberg generators $a,b$ by concrete
operators (differential, finite-difference, discrete) provided
that the vacuum remains unchanged leaves any polynomial operator
in $a,b$ isospectral. Now let us study another `face' of harmonic
oscillator by substituting the realization (\ref{e.3.2}) in
(\ref{e.2.6}). Finally, we obtain
\begin{equation}
\label{e.3.3}
h_d(y, D_{\pm})=
\frac{4}{\de}[y+\de(p+\frac{1}{2})]D_+ - 4(1 +\frac{1}
{\de})yD_- \ .
\end{equation}
Thus, in the realization by finite-difference operators the
corresponding spectral problem can be defined as
\[
\frac{4}{\de^2}\big[y+\de(p+\frac{1}{2})\big]\phi(y+\de)
- \frac{4}{\de}\big[(1+\frac{2}{\de})y+p+\frac{1}{2}\big]\phi(y)
+ \frac{4}{\de}(1+\frac{1}{\de})y\phi(y-\de)
\]
\begin{equation}
\label{e.3.4}
= E \phi(y) \ .
\end{equation}

The operator $h_d(y, D_{\pm})$ is a non-local, three-point,
finite-difference operator. It is illustrated by Fig.1.
\vskip .5cm
\noindent
\unitlength.8pt
\begin{picture}(400,50)(-10,-20)
\linethickness{1.2pt}
\put(60,10){\line(1,0){250}}
\put(120,10){\circle*{5}}
\put(180,10){\circle*{5}}
\put(240,10){\circle*{5}}
\put(90,-15){$\phi(y-\de)$}
\put(170,-15){$\phi(y)$}
\put(230,-15){$\phi(y+\de)$}
\end{picture}
\begin{center}
Fig. 1. \ Graphical representation of the operator (\ref{e.3.3})
\end{center}

In general, the function $\phi(y)$ in the rhs of (\ref{e.3.4})
can be replaced by $\phi(y+\de)$ or $\phi(y-\de)$, or by a linear
combination of $\phi(y\pm\de), \phi(y)$. It does not change the
statement that (\ref{e.3.4}) has infinitely-many polynomial
eigenfunctions.

It is easy to check that
\[
[ ye^{-\de \pa_y}]^n {\it I}\ =\ y^{(n)} {\it I}\ .
\]
where $y^{(n+1)}=y(y-\de)\ldots (y-n\de)$ is a so-called {\it
quasi-monomial} and $I$ is the identity operator. Using this
relation one can show that the eigenfunctions of (\ref{e.3.3})
remain polynomials and furthermore the solutions of (17) are
given by
\begin{equation}
\label{e.3.5}
 {\hat L}_n^{(p+\frac{1}{2})}(y,\de)=\sum_{\ell=0}^n
 a_{\ell}^{(p+\frac{1}{2})} y^{(\ell)}\ ,
\end{equation}
where $a_{\ell}^{(p+\frac{1}{2})}$ are the coefficients in the
expansion of the Laguerre polynomials, $L_n^{(p+\frac{1}{2})}(y)=
\sum_{\ell=0}^n a_{\ell}^{(p+\frac{1}{2})} y^{\ell}$. We call
these polynomials the {\it modified associated Laguerre
polynomials}. Simultaneously, the eigenvalues of the equation
(\ref{e.3.3}) remain equal to $(-4n), n=0,1,2\ldots$ and they are
the same as the eigenvalues of the harmonic oscillator problem
(1), (6) and (13). Thus, one can say that the operator
(\ref{e.3.3}) defines a {\it finite-difference form of harmonic
oscillator} Hamiltonian.

A natural question can be posed about the most general
second-order linear differential operator, which (i) has
infinitely-many polynomial eigenfunctions and (ii) is isospectral
to the harmonic oscillator (1). Following the Theorem
\cite{Turbiner:1994} one can show that this operator has a form
\[
h_g(y,\pa_y)\ =\ 4 (AJ^0+BJ^-) J^- - 4 J^0 + 4(p+\frac{1}{2})
 C J^-\ =
\]
\begin{equation}
\label{e4.1}
4(Ay+B)\pa_y^2 - 4[y-(p-\frac{1}{2})C]\pa_y\
\end{equation}
where $A,B,C$ are arbitrary constants and the generators
(\ref{e.2.3}) are realized by differential operators
(\ref{e.3.1}). However, by a linear change of variable, $y
\rar \al y + \be$ the operator (\ref{e4.1}) is transformed to
(5). Thus, without loss of generality we can put $A=\al=1$ and
also $C=1$. The eigenfunctions of (\ref{e4.1}) remain the
Laguerre polynomials but of a shifted argument,
$L_n^{(p-\frac{1}{2})} (y +\be)$. It leads to a statement that
among the second-order differential operators there exist no
non-trivial isospectral deformation of the harmonic oscillator
potential preserving polynomiality of the eigenfunctions.

The operator (\ref{e4.1}) can be rewritten in the Fock space
formalism by using (\ref{e.3.1})
\begin{equation}
\label{e4.2}
h_g(b,a)\ =\ 4(b+B)a^2 - 4[b-(p-\frac{1}{2})]a\ .
\end{equation}
It is evident that the operator (\ref{e4.2}) is the most general
second order polynomial in $a$, which is isospectral to
(\ref{e.2.6}) and also preserves the space of polynomials
(\ref{e.2.5}). By substitution (\ref{e.3.2}) into the operator
(\ref{e4.2}) it becomes transformed into a finite-difference
operator
\begin{equation}
\label{e.4.3}
h_g(y,D_{\pm})\ =\ 4BD_+^{2}+
\frac{4}{\de}[y+\de (p+\frac{1}{2})]D_+ -4(1 +\frac{1}{\de})yD_-
\end{equation}
(cf. (\ref{e.3.3})). Thus, in the realization by
finite-difference operators the corresponding spectral problem
can be defined as
\[
\frac{4B}{\de^2}\phi(y+2\de)+
\frac{4}{\de^2}\big[y-2B +\de (p+\frac{1}{2})\big]\phi(y+\de)
\]
\begin{equation}
\label{e.4.4}
 - \frac{4}{\de}\big[(1+\frac{2}{\de})y-\frac{B}{\de}+
 (p+\frac{1}{2})\big]\phi(y)
 + \frac{4}{\de}(1+\frac{1}{\de})y\phi(y-\de)\ =\ E \phi(y)\ ,
\end{equation}
and it has infinitely-many polynomial eigenfunctions.

The operator $h_d(y,D_{\pm})$ now becomes the four-point
finite-difference operator, see Fig.2 .
\vskip .5cm
\noindent
\unitlength.8pt
\begin{picture}(400,50)(-10,-20)
\linethickness{1.2pt}
\put(60,10){\line(1,0){300}}
\put(120,10){\circle*{5}}
\put(180,10){\circle*{5}}
\put(240,10){\circle*{5}}
\put(300,10){\circle*{5}}
\put(90,-15){$\phi(y-\de)$}
\put(170,-15){$\phi(y)$}
\put(220,-15){$\phi(y+\de)$}
\put(290,-15){$\phi(y+2\de)$}
\end{picture}
\begin{center}
Fig. 2. \ Graphical representation of the operator (\ref{e.4.3})
\end{center}

It is quite surprising that a linear shift of variable $y$ in
differential operator (\ref{e.3.3}) (which gives nothing
non-trivial, see discussion above) leads to occurrence of the
extra point in their isospectral finite-difference counterpart.

Now it is time to ask what would happen if in the expressions
(13),(20) the operators $a,b$ are not the generators of the
Heisenberg algebra (\ref{e.2.1}) but the generators of the
$q-$deformed Heisenberg algebra
\begin{equation}
\label{e.5.1}
             [a,b]_q \equiv ab  -  qba \ =\ 1,
\end{equation}
where $q$ is a parameter. Following the Theorem proved in
\cite{Turbiner:1994}, one can demonstrate that within the
$q-$deformed Fock space built on using $q-$deformed Heisenberg
algebra (23) there exists the flag of linear spaces of
polynomials in $b$, $\cal{P}$ (see (12)), which is preserved by
the operators (13),(20). By a simple calculation one can find the
eigenvalues of (13),(20)
\begin{equation}
\label{e.5.2}
             E_n^{(q)}\ =\ -4\{ n \}\ ,\ n=0,1,\ldots\ ,
\end{equation}
where
\[
 \{ n \}\ =\frac{1-q^n}{1-q}\ ,
\]
is a so-called $q-$number and $\{ n \}\rar n$, if $q \rar 1$. If
the parameter $q$ the spectra of (13),(20) are real.

The algebra (\ref{e.5.1}) has a realization in terms of discrete
operators (see, for example, \cite{Turbiner:1997})
\begin{equation}
\label{e.5.3}
a\ =\ {\cal D}_q ,\quad b\ =\ y \ ,
\end{equation}
where
\[
{\cal D}_q f(y) = \frac{f(qy) - f(y)}{y(1-q)}\ .
\]
This realization has a property that the vacuum remains the same
as well as for the cases (\ref{e.3.1})-(\ref{e.3.2}) and it can
be written as $|0>\
=\ 1$. Now we can substitute (\ref{e.5.3}) in (13) the following
operator emerges
\begin{equation}
\label{e.5.4}
h_q(y,{\cal D}_q)\ =\ 4\tilde J^0 \tilde J^- - 4 \tilde J^0 +
4(p+\frac{1}{2}) \tilde J^-
\ =\ 4y {\cal D}_q^2 - 4(y-p-\frac{1}{2}) {\cal D}_q\ ,
\end{equation}
where the generators $\tilde J^0=ba,\ \tilde J^-=a$ have the same
functional form as (\ref{e.2.3}) but obey the $q-$ deformed
commutation relation
\[
[\tilde J^0,\tilde J^-]_{1/q} \ \equiv\ \tilde J^0 \tilde J^- -
\frac{1}{q}\tilde J^0\tilde J^-\ =\ -\tilde J^-\ ,
\]
forming the $q-$deformed Borel subalgebra $b(2)_{q}$ of the
$q-$deformed algebra $sl(2)_q$.

The operator $h_q(y, D_q)$ is a non-local, three-point, discrete,
dilatation-invariant operator illustrated by Fig.3.

\vskip .5cm
\noindent
\unitlength.8pt
\begin{picture}(400,50)(-10,-20)
\linethickness{1.2pt}
\put(60,10){\line(1,0){310}}
\put(120,10){\circle*{5}}
\put(180,10){\circle*{5}}
\put(300,10){\circle*{5}}
\put(110,-15){$\phi(y)$}
\put(170,-15){$\phi(qy)$}
\put(280,-15){$\phi(q^2y)$}
\end{picture}
\begin{center}
Fig. 3. \ Graphical representation of the operator (\ref{e.3.3})
\end{center}

\vskip .5cm

The spectral problem for the operator (\ref{e.5.4}) has a form
\[
4\frac{\phi(q^2y)}{yq(q-1)^2} -4
\frac{1+q+(y-p-\frac{1}{2})q(1-q)}{yq(q-1)^2}\phi(qy)+
4
\frac{1+(y-p-\frac{1}{2})(1-q)}{y(q-1)^2}\phi(y)
\]
\begin{equation}
\label{e.5.5}
\ =\ E^{(q)} \phi(y) \ .
\end{equation}
or, the rhs can be taken as
\begin{equation}
\label{e.5.6}
\ =\ E^{(q)} \phi(qy) \ .
\end{equation}
or as
\begin{equation}
\label{e.5.7}
\ =\ E^{(q)} \phi(q^2y) \ .
\end{equation}

 If for the case (\ref{e.5.5}) the eigenvalues are given by
(\ref{e.5.2}) while for (\ref{e.5.6}), (\ref{e.5.7}) the
eigenvalues are equal to
\begin{equation}
\label{e.5.8}
             E_n^{(q)}\ =\ -4 q^n \{ n \}\ ,\ n=0,1,\ldots
\end{equation}
\begin{equation}
\label{e.5.9}
             E_n^{(q)}\ =\ -4 q^{2n} \{ n \}\ ,\ n=0,1,\ldots
\end{equation}
correspondingly, while in the limit $q \rar 1$ all three
expressions coincide. The spectral problems
(\ref{e.5.5})--(\ref{e.5.7}) can be considered as a possible
definition of a $q-$deformed harmonic oscillator. In the
literature it is known many other definitions of the $q-$deformed
harmonic oscillator. Such a situation reflects an existence of an
ambiguity appearing when a $q-$ deformation is performed
\footnote{For example, any term in non-deformed expression can be
modified by multipliers of the type $q^a$ and it can be added
extra terms with vanishing coefficients in the limit $q\rar 1$
like $(1-q)^b$.} and absence of clear criteria, which can remove
or reduce this ambiguity. For instance, in the literature it is
exploited three types of the $q-$Laguerre polynomials (see, for
example, an excellent review \cite{Koekoek:1994}), but it is not
clear why other possible $q-$deformations of Laguerre polynomials
are not studied.

Substitution of (25) in (20) gives a slight modification of the
expressions (26)- (27). Unlike translation-invariant case it does
not lead to a change of the number of points in the operator (26)
as it happened for the operators (16) and (21).

\newpage
\def\href#1#2{#2}

\begingroup\raggedright\endgroup

\end{document}